\documentclass[twocolumn]{aastex62}

\usepackage[latin1]{inputenc}
\usepackage{tikz}
\usepackage{amssymb,mathtools}
\usetikzlibrary{shapes,arrows}
\tikzstyle{block} = [rectangle, draw, 
text width=15em, text centered, minimum height=2em]
\tikzstyle{arrow} = [thick,->,>=stealth]

\shorttitle{FPI turbulence in solar wind}
\shortauthors{Bandyopadhyay et al.}

\begin{document}
	
\title{Solar Wind Turbulence Studies using MMS Fast Plasma Investigation Data}

\author[0000-0002-6962-0959]{Riddhi Bandyopadhyay}
\affiliation{Department of Physics and Astronomy, University of Delaware, Newark, DE 19716, USA}
	
\author[0000-0001-8478-5797]{A. Chasapis}
\affiliation{Department of Physics and Astronomy, University of Delaware, Newark, DE 19716, USA}
	
\author[0000-0003-0602-8381]{R. Chhiber}
\affiliation{Department of Physics and Astronomy, University of Delaware, Newark, DE 19716, USA}

\author[0000-0003-0602-8381]{T.~N. Parashar}
\affiliation{Department of Physics and Astronomy, University of Delaware, Newark, DE 19716, USA}
\affiliation{Bartol Research Institute, University of Delaware, Newark, DE 19716, USA}

\author[0000-0002-2229-5618]{B.~A. Maruca}
\affiliation{Department of Physics and Astronomy, University of Delaware, Newark, DE 19716, USA}	
	
\author[0000-0001-7224-6024]{W.~H. Matthaeus}
\email{whm@udel.edu}
\affiliation{Department of Physics and Astronomy, University of Delaware, Newark, DE 19716, USA}
\affiliation{Bartol Research Institute, University of Delaware, Newark, DE 19716, USA}

\author{S.~J. Schwartz}
\email{}
\affiliation{Laboratory for Atmospheric and Space Physics, University of Colorado Boulder,  Boulder, Colorado, USA}

\author[0000-0002-5619-1577]{S. Eriksson}
\email{}
\affiliation{Laboratory for Atmospheric and Space Physics, University of Colorado Boulder,  Boulder, Colorado, USA}

\author{O. Le~Contel}
\affiliation{CNRS/Ecole Polytechnique/Sorbonne Universit\'e/Univ. Paris Sud/Observatoire de Paris, Paris, France}

\author{H. Breuillard}
\affiliation{Laboratoire de Physique des Plasmas, UMR7648, CNRS/Ecole Polytechnique/Sorbonne Univ./Univ. Paris Sud, Paris, France}

\author[0000-0003-0452-8403]{J.~L. Burch}
\affiliation{Southwest Research Institute, San Antonio, TX, USA}

\author[0000-0002-3150-1137]{T.~E. Moore} 
\affiliation{NASA Goddard Space Flight Center, Greenbelt, MD, USA}

\author[0000-0001-9249-3540]{C.~J. Pollock}
\affiliation{Denali Scientific, Fairbanks, Alaska, USA}

\author[0000-0001-8054-825X]{B.~L. Giles}
\affiliation{NASA Goddard Space Flight Center, Greenbelt, MD, USA}

\author{W.~R. Paterson}
\affiliation{NASA Goddard Space Flight Center, Greenbelt, MD, USA}

\author{J. Dorelli}
\affiliation{NASA Goddard Space Flight Center, Greenbelt, MD, USA}

\author[0000-0003-1304-4769]{D.~J. Gershman}
\affiliation{NASA Goddard Space Flight Center, Greenbelt, MD, USA}

\author{R.~B. Torbert}
\affiliation{University of New Hampshire, Durham, NH, USA}

\author[0000-0003-1639-8298]{C.~T. Russell}
\affiliation{University of California, Los Angeles, CA, USA}

\author[0000-0001-9839-1828]{R.~J. Strangeway}
\affiliation{University of California, Los Angeles, CA, USA}

	
	
\begin{abstract}
Studies of solar wind turbulence 
traditionally employ high-resolution magnetic field data, 
but high-resolution measurements 
of ion and electron moments have been possible only recently. 
We report the first turbulence studies of ion and electron 
velocity moments 
accumulated in pristine solar wind 
by the Fast Particle Investigation instrument onboard the 
Magnetospheric Multiscale (MMS) Mission. 
Use of these data is made possible by a novel implementation 
of a frequency domain Hampel filter, described herein.
After presenting procedures for processing of the data, 
we discuss statistical properties of solar wind turbulence
extending into the kinetic range. 
Magnetic field fluctuations dominate electron and ion velocity fluctuation 
spectra 
throughout the energy-containing and inertial ranges. 
However, a multi-spacecraft analysis indicates that at scales shorter 
than the ion-inertial length, electron velocity fluctuations become 
larger than ion velocity and magnetic field fluctuations. 
The kurtosis of ion velocity peaks around few ion-inertial 
lengths and returns to near gaussian value at sub-ion scales.
\end{abstract}



\section{Introduction}\label{sec:intro}
The solar wind is hot, diffuse, supersonic plasma flow from the Sun. The solar wind provides a natural laboratory for study of plasma turbulence~\citep{Tu1995SSR, Bruno2005LRSP}. 
In recent years, the nature of kinetic scale turbulence in the solar wind has been of considerable 
interest \citep{Bale2005PRL, Alexandrova2009PRL, Sahraoui2010PRL, Alexandrova2012ApJ, Salem2012ApJL, Roberts2017ApJ}. 
Availability of high time-resolution magnetometer instruments has prompted the community to carry out most of these studies using magnetic field data. Due to practical constraints, kinetic scale studies with velocity data remain rare. This factor is particularly inconvenient from the perspective of turbulence studies, since a full understanding of a turbulent plasma requires multi-scale information of both magnetic field and ion and electron distribution function moments. In fact, the equations of magnetohydrodynamics (MHD) become symmetric and physically revealing when written 
in terms of Elsasser variables (e.g., see~\cite{Politano1998GRL,Politano1998PRE}), 
thus calling for high-resolution measurements of both magnetic field and velocity field.

The Magnetospheric Multiscale (MMS) Mission was launched in 2015 with the primary objective of studying magnetic reconnection in the terrestrial magnetopause and magnetotail~\citep{Burch2016SSR}. The very high time resolution of the plasma instruments  onboard MMS along with the availability of simultaneous measurements from four spacecraft, separated by sub-ion scale distances, make MMS observations an excellent case for study of inertial and kinetic scale turbulence in near-Earth space. We note in passing that there have been numerous studies of turbulence in the magnetospheric environment, using Cluster~\citep{Retino2007NaturePh,Sundkvist2007PRL,Chasapis2015ApJL}, MMS~\citep{Yordanova2016GRL,Stawarz2016JGR,Chen2017ApJ,Chasapis2017ApJ,Roberts2018GRL} as well as some earlier single-spacecraft studies~\citep{Borovsky1997JPP}. Here, we carry out analogous studies in the distinct turbulence environment of the solar wind using high resolution MMS data. 

Near the end of 2017, a few reasonably long $(\approx 1~\mathrm{hour})$ MMS burst mode measurement intervals of the pristine solar wind, outside the bow shock, were made available. We present here the first study of high resolution 
statistical 
turbulence properties of the solar wind using simultaneous measurement of ion and electron moments, and magnetic field data from MMS. In particular, we take advantage of the FPI instrument which provides accurate measurements of ion and electron distribution moments with unprecedented time resolution. 
In order to exploit the advantages of the MMS/FPI instrumentation, we find it to be 
necessary to 
carry out some technical refinements in the form of filtering, to render the computed 
FPI moments to be of suitable quality for this study.
Additionally, MMS multi-spacecraft 
capability allows us to directly observe sub-ion scale turbulence without the 
need of assuming Taylor's frozen-in hypothesis.
This affords an unusual opportunity to directly compare single spacecraft Taylor-hypothesis 
correlations with two-spacecraft direct correlation measurements. 
We note that in ideal circumstances, when the turbulence is isotropic and the Taylor hypothesis is valid, the 2-point, single-time measurements and the single-point, time-lagged measurements are theoretically equivalent.  Various factors cause this relationship to breakdown, in part, or even entirely~\citep{Sahraoui2010JGR}. For example, when the turbulence is anisotropic relative to the radial (or other) direction, spacecraft separated in the direction transverse to radial may not agree well with (radial) frozen-in flow correlations. Finite integration times for individual data points also can blur the interpretation. These issues will be discussed further.

\section{Overview}\label{sec:data}
\begin{figure*}
	\begin{center}
		\includegraphics[scale=0.6]{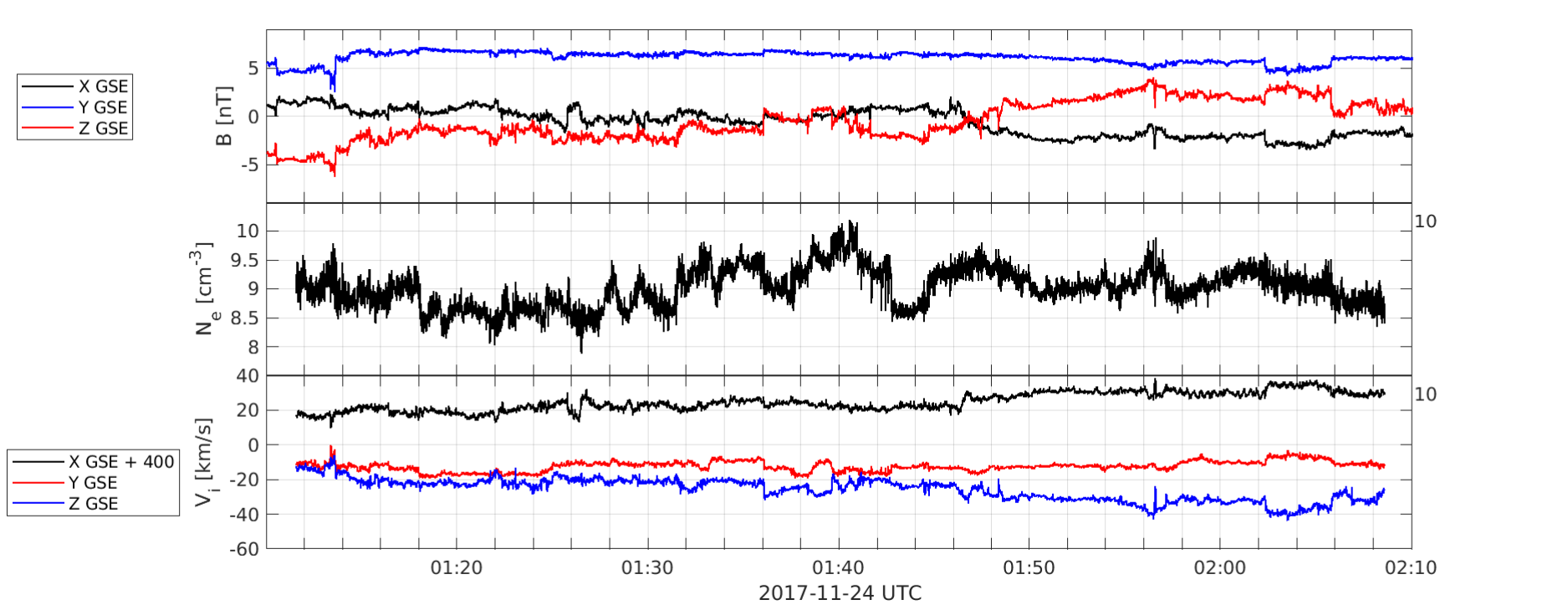}
		\caption{
			Overview of the MMS observations in solar wind turbulence selected for this study. The data shown is from the FGM and FPI instruments on-board the MMS1 spacecraft. Top panel shows the magnetic field measurements, middle panel shows the electron density, bottom panel shows the ion velocity in GSE coordinates. The X component of velocity has been shifted by $400~\mathrm{km~s^{-1}}$. The FPI data have been processed by the procedure described in the text.}
		\label{fig:overview}
	\end{center}
\end{figure*}

In late 2017, the MMS
apogee was at $\sim 27~R_{\mathrm{E}}$ on Earth's day side of the magnetosphere and outside the ion foreshock region. This orbit allowed the spacecraft to sample the pristine solar wind, outside the Earth's magnetosheath, for extended periods of time. 
It should be noted that, for statistical studies of turbulence in the solar wind, relatively long continuous intervals are needed 
of duration corresponding to 
at least a few times the typical correlation scale. 
Here, we focus on an interval of approximately one hour length on 2017 November 24 from 01:10:03 to 02:10:03 UT. Figure~\ref{fig:overview} shows an overview of this MMS data interval.
 
The burst mode magnetic and electric field data were provided by the FIELDS instrument suite~\citep{Torbert2016SSR}. Specifically, the flux-gate magnetometer (FGM) measured the vector magnetic field at a resolution of $128~\mathrm{Hz}$~\citep{Russell2016SSR}. The FPI instrument~\citep{Pollock2016SSR} measures the ion and electron distribution functions and calculates the moments of those distributions with a cadence of $150~\mathrm{ms}$ and $30~\mathrm{ms}$ respectively. 

An overview of the relevant parameters for the interval of interest can be seen in Table~\ref{tab:overview} which reports plasma parameters: mean magnetic field strength $|\langle \mathbf{B} \rangle|$, the rms fluctuation value of the magnetic field $\delta B = \sqrt{\langle |\mathbf{B}(t) - \langle \mathbf{B} \rangle|^2 \rangle}$, ion inertial length $d_\mathrm{i} $, electron inertial length $d_\mathrm{e}$, the solar wind speed $V_{\mathrm{SW}}$ and the proton plasma beta $\beta_p$ are reported.
\begin{table}[ht!]
	\caption{Overview of the selected interval} 
	\label{tab:overview}
	\begin{center}
		\begin{tabular}{c c c c c c c}
			\hline \hline
			$|\langle \mathbf{B} \rangle|$ & $\delta B /|\langle \mathbf{B} \rangle|$ 
			& $\langle n_e \rangle$  
			& $d_{\mathrm{i}}$ & $d_{\mathrm{e}}$ 
			&$V_{\mathrm{SW}}$ 
			& $\beta_p$ \\
			\colhead{($\rm nT$)} & \colhead{} & \colhead{(${\rm cm^{-3}}$)} & \colhead{($\rm km$)} & \colhead{($\rm km$)} & \colhead{($\rm km~s^{-1}$)} & \colhead{} \\
			\hline
			6.6 & 0.4 & 8.6 & 75.9 & 1.8 & 377 &1.3\\
			\hline
		\end{tabular}
	\end{center}
	\tablecomments{Data obtained from MMS1 on 2017 November 24 from 01:10:03 to 02:10:03 UT.}
\end{table}

Due to the difficulties of FPI measurements in the solar wind, as discussed earlier, some systematic uncertainties may exist in the plasma moments. These uncertainties become more prominent in higher moments like temperature. We have cross-checked the parameters of Table~\ref{tab:overview} with Wind FC and MFI data from around 00:40:00 UTC (i.e.,$\sim1~\mathrm{hour}$ before the midpoint of the period used in this study). While the density, velocity, and magnetic field values agree very well, significant deviations were present in the temperature and consequently the proton beta estimates by the FPI instrument and the Wind estimates. The Wind measurements resulted in the proton beta of $\beta_{p} = 0.43$. Given the limitations of FPI instruments in the solar wind, the Wind measurements of temperature and proton beta are more reliable.

\section{FPI moments in the solar wind}\label{sec:filter}
\begin{figure}
\begin{center}
\includegraphics[scale=0.9]{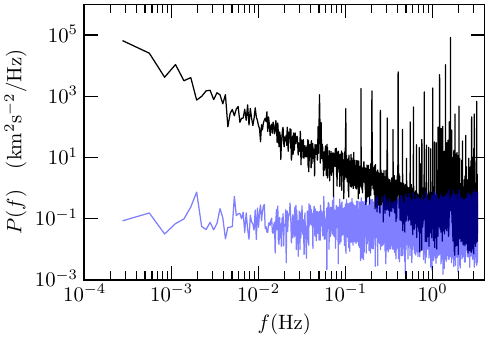}
\caption{Power spectrum of ion velocity from unfiltered level 2 burst mode data shown in black. The noise floor is shown in translucent blue. The satellite spin frequency is 0.05 Hz which corresponds to the first spike in the spectrum.}
\label{fig:Ef_vi}
\end{center}
\end{figure}

Contrary to the magnetosheath and the magnetopause, a number of issues arise when using FPI data in the solar wind. Many of these problems arise because the FPI detectors are, optimized for magnetospheric response. However, when operating in the solar wind, almost all particles enter in just a few angular channels. This introduces problems such as periodicities (harmonics of the spacecraft rotation) that are due to, for example, the solar wind particles beam crossing between the detectors as the spacecraft rotates. The main implication for the present study is the occurrence of large-amplitude fluctuations in the time series of the plasma moments. These systematic features
correspond to numerous
very narrow, large amplitude peaks in the frequency spectrum as seen in Figure~\ref{fig:Ef_vi}.
They are present in several burst resolution solar wind intervals, including the one used for the results presented here.
These artifacts are thought to be associated, in part, 
with gaps in the angular coverage of the particle 
distribution functions. It is worth noting here that the first spike in the spectrum is due to the satellite spintone at 0.05 Hz.

An in-depth analysis of the FPI spectrometers' response in the solar wind is beyond the scope of this study.
We apply a specially designed filter, based on time-series analysis, to mitigate the effect of these instrumental artifacts to proceed with this study.

For the present work, an algorithm was developed
to filter out the artifacts depicted in Figure \ref{fig:Ef_vi}. 
This kind of scheme is feasible since, although they have large amplitude, for most of the cases, 
these features 
are very narrow in frequency space as seen in Figure~\ref{fig:Ef_vi}. 
An exception is a broad peak at $\simeq 1.625~\mathrm{Hz}$, which persists in the filtered signal as well, as shown later. 
It should be noted that this frequency is close to the $1/32$ of the spin period $(\approx 20~\mathrm{s})$ of the spacecraft. Given that the particle distributions are measured in 32 azimuthal bins on the spin plane of the spacecraft, this suggests that the broad peak near $1.625~\mathrm{Hz}$ may be  instrumental in origin. 
Since we employ a low-pass filter to remove the part of the spectrum where the signal is dominated by the noise, this feature does not impact the present study.

\begin{figure}
	\begin{center}
		\includegraphics[scale=0.7]{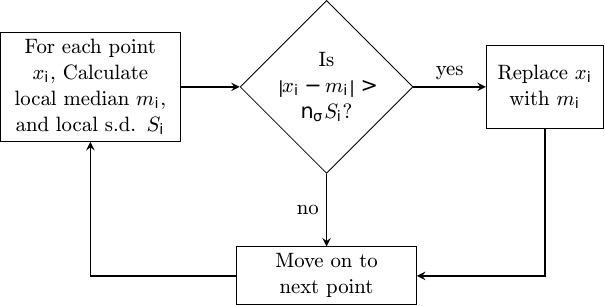}
		\caption{Schematic diagram of working mechanism of the Hampel filter. s.d. $\equiv$  standard deviation.}
		\label{fig:hampel}
	\end{center}
\end{figure}

\textit{Implementation of a spectral Hampel filter and time series reconstruction:} 
To eliminate these spikes, we adapt a decision-based filter known as 
\textit{Hampel filter}, a well-known technique in the signal-processing community for 
removing outliers from a signal~\citep{Davies1993JASA,Liu2004CCE,Pearson2016JASP}. 
We apply the filter in frequency spectra domain to remove 
the spikes, while maintaining the original phase information.
We then reconstruct 
the time series via an inverse Fourier transform.\footnote{This procedure differs from a typical application of the Hampel filter in which the outlying points in the time domain are trimmed to have amplitude similar to their neighbors. We adapt this idea here to frequency space.} 
The basic working mechanism of the filter is outlined in Figure~\ref{fig:hampel}. 
It is distinct from a window average or median filter which has only one tuning parameter: the window width. The Hampel filter works as follows: for each point, $x_{\mathrm{i}}$ in a given series, 
the local median, $m_{\mathrm{i}}$ and the local standard deviation, $S_{\mathrm{i}}$ are calculated over a window of predetermined size (say, $k$) around that point. If absolute difference of the value of that point and the local median, $|x_{\mathrm{i}}-m_{\mathrm{i}}|$ is above a threshold defined as $n_{\sigma}$, times the local 
standard deviation, the value is replaced by the median. 
If not, the algorithm leaves the current point unchanged and proceeds to the next point. Mathematically, the filter's response can be described as:
\begin{equation}\label{eq:hampel}
y_{\mathrm{i}} =
\begin{cases}
x_{\mathrm{i}} & \text{if $|x_{\mathrm{i}}-m_{\mathrm{i}}|\le n_{\sigma} S_{\mathrm{i}}$,} \\
m_{\mathrm{i}} & \text{if $|x_{\mathrm{i}}-m_{\mathrm{i}}|> n_{\sigma} S_{\mathrm{i}}$,}
\end{cases}
\end{equation}
where $x_{\mathrm{i}}$ is the particular point under consideration, $m_{\mathrm{i}}$ is the local median value from the moving window,
\begin{eqnarray}
m_{\mathrm{i}} = \mathrm{median}(x_{\mathrm{i-k}},x_{\mathrm{i-k+1}},\cdots,x_{\mathrm{i+k-1}},x_{\mathrm{i+k}})\label{eq:median},
\end{eqnarray}
and $S_{\mathrm{i}}$ is the local standard deviation estimated as,
\begin{eqnarray}
S_{\mathrm{i}} = \kappa \times \mathrm{median} (|x_{\mathrm{i-k}}-m_{\mathrm{i}}|,\cdots,|x_{\mathrm{i+k}}-m_{\mathrm{i}}|)\label{eq:std},
\end{eqnarray}
where $\kappa = 1/\sqrt{2}~\mathrm{erfc}^{-1}(1/2) \approx 1.4826$ is an estimate of the standard deviation for Gaussian noise. In the usual cases, the filter reduces to a standard median filter when the threshold parameter $n_{\sigma}$ is set to zero, and becomes an identity operator as $n_{\sigma} \rightarrow \infty$.
	
In our case however, the outliers exist in frequency space. Therefore, we apply the filter to the Fourier transform of the signal. Localized peaks with extreme values are replaced by the local 
median value, 
while the phase information is conserved. Therefore, we are able to perform an inverse Fourier 
transform which allows us to reconstruct
the original signal without the undesired artifacts. This approach was found to perform significantly better than the use of notch filters, due to the fact that the peaks were very localized in frequency space.

We apply this technique to  power spectral density (PSD) of relevant quantities with a local window of $101$ points in frequency domain, choosing a threshold value of $n_{\sigma} = 3$. The window length of 101 points, corresponding to a frequency of $0.0028~\mathrm{Hz}$, is heuristically determined here. For this particular interval, this choice of parameters helps to eliminate most of the spikes in the spectra while leaving the underlying signal unchanged. For other data intervals, the parameters may need to be changed. Applying the filter effectively removes the undesired spectral features which we believe are associated with instrumental artifacts induced by the operation of FPI instrument in the solar wind. A diagnostic of the filter algorithm employed and an evaluation of its performance is detailed in the Appendix.

\begin{figure}
	\begin{center}
		\includegraphics[scale=0.8]{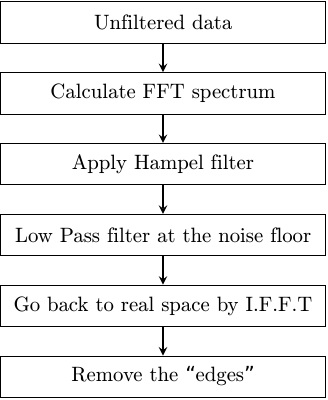}
		\caption{Schematic diagram of required steps for removing instrumental artifact from the signal. FFT $\equiv$ Fast Fourier Transform, IFFT $\equiv$  Inverse Fast Fourier Transform.}
		\label{fig:flow}
	\end{center}
\end{figure}

We calculated the noise floor as described in~\cite{Gershman2018PoP} and determined that the ion velocity spectrum becomes noise dominant near $\approx 1~\mathrm{Hz}$ (see Figure~\ref{fig:Ef_vi}). Therefore, we  applied a low-pass filter to the signal at the noise floor and set all higher frequencies to zero. Finally, after applying the Hampel filter, edge effects create some oscillations at the start and end of the time series. 
So we have removed from the analysis the first and last 
$500$ 
points 
of the time-series. It is worth mentioning here that the interval of these ``edge effects" depend on the total length of the time series. For shorter intervals the edge oscillations are confined to narrower regions. 
Figure~\ref{fig:flow} outlines schematically all the steps required to obtain the final signal sufficiently free of instrumental artifacts.

\section{Turbulence Results}\label{sec:results}
\begin{figure}
\begin{center}
\includegraphics[scale=1.0]{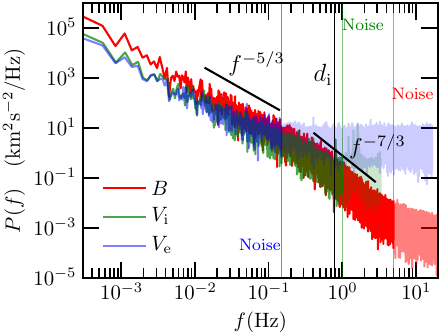}
\caption{FFT spectra of magnetic field, Hampel-filtered ion and electron velocity field as measured by MMS1. The  colored vertical lines represent where the signal to noise ratios become smaller than about $5$, after which the rest of each spectrum has been plotted with high transparency The black solid line represent $k d_\mathrm{i}=1$, where $k \simeq 2 \pi f/|\langle \mathbf{V} \rangle|$.}
\label{fig:spectrum}
\end{center}
\end{figure}

In Figure~\ref{fig:spectrum}, we show the spectral power density (trace of the spectral density matrix) of magnetic field, ion velocity, and electron velocity. The magnetic field has been converted to Alfv\'en units $; \mathbf{B} \rightarrow \mathbf{B}/\sqrt{\mu_0 m_p n_0}$, where $\mu_0$ is the vacuum permeability, $m_p$ is the proton mass and $n_0=\langle n_e\rangle$ is the mean number density of electrons $\approx \langle n_p\rangle$, the average number density of protons, assuming quasi-neutrality. While the electron density tends to be more reliable that the ion density for MMS observations, in this interval both were found to be within $\approx 5 \%$ of each other and in agreement with the corresponding values obtained from Wind data. 

The spectral noise floors were determined by the procedure described in~\cite{Gershman2018PoP}. 
Figure~\ref{fig:spectrum} shows that for the electron velocity measurements
the noise becomes significant near $\approx 0.15~\mathrm{Hz}$, while for the ion velocity  this occurs at $\approx 1~\mathrm{Hz}$. Note that, for burst mode MMS data in the Earth's magnetosheath, usually ion velocity spectra hits the noise floor before electron velocity  (see~\cite{Gershman2018PoP}). So one can conclude that for MMS measurements in the pristine solar wind, the ion velocity effectively covers a better frequency range than the electron velocity while the situation is reversed for the magnetosheath. The magnetic field spectrum obtained by FGM crosses the instrumental noise floor near $\sim 5~\mathrm{Hz}$~\citep{Russell2016SSR}.

\begin{figure}
\begin{center}
\includegraphics[scale=1.0]{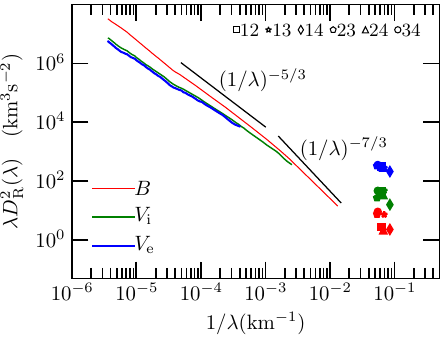}
\caption{Equivalent spectra of magnetic, ion velocity, and electron velocity. The ion and electron velocity fields have been processed as described in Sec~\ref{sec:filter}. The continuous lines in color red, green and blue represent Taylor hypothesis based magnetic, ion-velocity, and electron velocity, in order of increasing thickness respectively. The symbols are multi-spacecraft values.}
\label{fig:equiv}
\end{center}
\end{figure}

An approximate Kolmogorov scaling $\sim f^{-5/3}$ can be seen in the inertial range for all three variables, followed by a steepening in both magnetic field and ion velocity field spectra in Figure~\ref{fig:spectrum}. To investigate the behavior of each quantity more clearly, we proceed to calculate the ``equivalent spectra," as shown in Figure~\ref{fig:equiv}. 
For single-spacecraft analysis the time increments of the velocity (or any other vector field) are defined as
\begin{equation}
\Delta V_i^{t} (\tau) = V_i(t+\tau) - V_i(t)\label{eq:1sc},
\end{equation}
where $i$ represents $R$(radial)=$\mathrm{X_{GSE}}$, $T$(tangential)=$\mathrm{Y_{GSE}}$ or $N$(normal)=$\mathrm{Z_{GSE}}$ component of the velocity field, referring to the \textit{Geocentric Solar Ecliptic} (GSE) coordinate~\citep{Franz2002PSS}. For the multi-spacecraft technique, we define the increment of the $i^{\mathrm{th}}$ component as
\begin{eqnarray}
\Delta V_i^{ab}(t) = V_i^{a}(t) - V_i^{b}(t)\label{eq:2sc},
\end{eqnarray}
where $t$ is the time of measurement and the indices $a, b$ represent any of the four MMS spacecraft. The increment in this case corresponds to a spatial separation or lag of $\mathbf{r}_{ab} = \mathbf{x}_{a}-\mathbf{x}_{b}$, where $\mathbf{x}_{a}$ and $\mathbf{x}_{b}$ are the positions of the two spacecraft. This enables one to evaluate actual spatial increment without the need 
to invoke the Taylor hypothesis. The second-order longitudinal structure function of the vector field is defined as,
\begin{eqnarray}
D^{(2)}_{\mathrm{R}}(\lambda) = \langle [\Delta V^{t}_{\mathrm{R}}(\lambda)]^{2} \rangle \label{eq:long1sc},
\end{eqnarray}
where $\langle \cdots \rangle$ represents an average over the full time series. The time-scales of the non-linear processes we are interested in are much larger than the time scales of the flow of the plasma past the spacecraft. For single-spacecraft solar wind data, therefore, it is assumed that the plasma is frozen-in and the available time series data is equivalent to a one dimensional spatial data along the plasma system. Operationally, to a good approximation, the associated spatial increments $\lambda$ are given by the Taylor hypothesis~\citep{Taylor1938PRSLA} as
\begin{eqnarray}
\lambda \hat R = - V_{\mathrm{sw}}\tau \hat R\label{eq:taylor},
\end{eqnarray} 
where $\hat R$ is the unit vector along the radial direction and $\tau$ is the time increment. The multi-spacecraft version of the calculation of structure function reads similarly
\begin{eqnarray}
D^{(2)}_{\mathrm{R}}(r_{ab}) = \langle [\Delta V^{ab}_{\mathrm{R}}(t)]^{2} \rangle \label{eq:long2sc}.
\end{eqnarray}
The quantity $S^{2}(\lambda)=\lambda D^{2}_{\mathrm{R}}(\lambda)$ then behaves as an \textit{equivalent spectrum} viewed as a function of an effective wavenumber $k^{\ast}=1/\lambda$. If the omnidirectional wavenumber spectra exhibits a power-law scaling $E(k) \sim k^{-\alpha}$ for a range of bandwidth, one expects the equivalent spectrum to scale similarly $S^{(2)}(\lambda) \sim (1/\lambda)^{-5/3}$ for a similar range of $k^{\ast}=1/\lambda$. Equivalent spectra are useful for analyzing statistics of data sets which are relatively limited in time duration as the present one~\citep{MoninYaglom-vol1,MoninYaglom-vol2}. Figure~\ref{fig:equiv} illustrates the equivalent spectra of magnetic, ion-velocity, and electron-velocity field, computed from the longitudinal second-order structure functions. The ion and electron velocities have been processed as described in Sec~\ref{sec:filter}. Only the parts of the spectra, unaffected by the instrumental noise, are plotted. The ion and electron velocities follow each other almost perfectly in the large-scale and inertial-scale ranges. A Kolmogorov scaling is clearly seen for the magnetic field equivalent spectrum in the inertial range. The ion and electron velocity spectra show a shallower scaling, as was observed in \cite{Podesta2006JGR}. 

Due to the effect of noise, we can not extend the electron-velocity spectrum further into the kinetic range, so we adopt a multi-spacecraft technique~\citep{Horbury2000Cluster}.For the present interval, the average separation of the four MMS spacecraft is about $15.6~\mathrm{km}$. These separation lengths are smaller than ion-inertial length $d_{\mathrm{i}}$, but still larger than the electron-inertial length $d_{\mathrm{e}}$ (see Table~\ref{tab:overview}). 

The reader should note that the time for convection of the solar wind over a distance comparable to the spacecraft separation $(\sim 20~\mathrm{km})$, assuming a $\sim 400~\mathrm{km}~\mathrm{s^{-1}}$ wind speed, is about $0.05~\mathrm{s}$. This advection time is less than the integration time for an ion Velocioty Distribution Function $(150~\mathrm{ms})$, but less than the integration time for the electron distributions $(30~\mathrm{ms})$. Therefore, the blurring of the spatial information due to finite integration time is a greater concern for the ion distributions. Even though the spacecraft pairs are not aligned in the radial direction (flow direction) for this interval, this influence causes us to have most confidence in the magnetic-field two-spacecraft structure functions, and least confidence in the ion-velocity structure functions. 

Further, the Taylor hypothesis needs to be modified from its single-spacecraft version $(V_{\mathrm{SW}}\gg V_{\mathrm{A}})$ for multi-spacecraft analysis~\citep{Horbury2000Cluster,Osman2007ApJL,Osman2009JGR,Chen2010PRL}. For two spacecraft, separated by a displacement $\mathrm{\mathbf{d}}_{1,2}$, the frozen-in condition is satisfied when
\begin{eqnarray}
\frac{V_{\mathrm{SW}} \Delta t}{|\mathrm{\mathbf{d}}_{1,2}- \mathbf{V}_{\mathrm{SW}}\Delta t|} \frac{V_{\mathrm{A}}}{V_{\mathrm{SW}}} \ll 1 \label{eq:taylor-multi},
\end{eqnarray}	
where $V_{\mathrm{A}}$is the Alfv\'en speed and $\Delta t$ is the sampling time of each spacecraft. We checked the criterion equation~(\ref{eq:taylor-multi}) for all the measurements used here. The left side of equation~(\ref{eq:taylor-multi}) always remains below $0.12$, so the condition imposed by equation~(\ref{eq:taylor-multi}) is well satisfied for the measurements employed here. 

The multi-spacecraft estimates reveal that the electron velocity exceeds the ion velocity and magnetic field at the spacecraft separation scales, which are deep inside the kinetic range. This kind of comparison is a significant advantage of the current approach in which we employ both single and multi-spacecraft techniques. Similar energy partitioning was found for low-beta plasma in the magnetosheath by~\cite{Gershman2018PoP}. Since plasma $\beta$ for the present interval is lower than one (according to the Wind measurements $\beta_{p}=0.43$) our results are qualitatively consistent with~\cite{Gershman2018PoP}.  

\begin{figure}
\begin{center}
\includegraphics[scale=1.1]{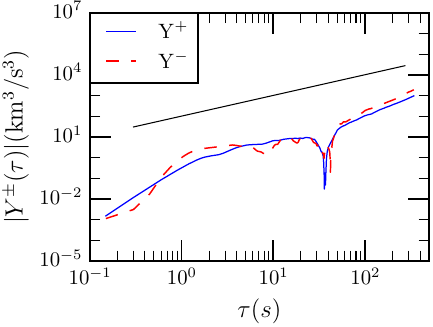}
\caption{Third-order mixed structure function. A linear scaling is shown for reference.}
\label{fig:3rd}
\end{center}
\end{figure}

We now proceed to calculate third-order statistics of the fields, related to turbulent dissipation of the plasma. The Kolmogorov-Yaglom law, generalized to isotropic MHD~\citep{Politano1998GRL,Politano1998PRE}, in the inertial range reads
\begin{eqnarray}
Y^{\pm}(r) = - (4/3) \epsilon^{\pm} r \label{eq:3rd},
\end{eqnarray}
where 
\begin{eqnarray}
Y^{\pm}(r)=\langle \mathbf{\hat{r}}\cdot \Delta \mathbf{Z}^{\mp}(\mathbf{x},\mathbf{r}) |\Delta \mathbf{Z}^{\pm}(\mathbf{x},\mathbf{r})|^2 \rangle \label{eq:ypm},
\end{eqnarray}
and
\begin{eqnarray}
\mathbf{Z}^{\pm} = \mathbf{V} \pm \mathbf{B}\label{eq:zpm}.
\end{eqnarray}
Here $\langle \cdots \rangle$ represents ensemble average, and $\mathbf{B}$ is in Alfv\'en unit. The variables $\Delta \mathbf{Z}^{\pm}(\mathbf{x},\mathbf{r})=\mathbf{Z}^{\pm}(\mathbf{x}+\mathbf{r})-\mathbf{Z}^{\pm}(\mathbf{x})$ are the \textit{increments} of the Elsasser variables along a spatial lag $\mathbf{r}$. The quantities $\epsilon^{+}$ and $\epsilon^{-}$ are the mean dissipation rates of corresponding Elsasser variables. The total (magnetic$+$kinetic) energy dissipation rate is given by
\begin{eqnarray}
\epsilon = \bigg( \frac{\epsilon^{+}+\epsilon^{-}}{2} \bigg)\label{eq:eps}.
\end{eqnarray}
In order to calculate the Elsasser variables (Eq.~\ref{eq:zpm}) from MMS data, the ion and magnetic-field measurements of each spacecraft were synchronized, and the latter were resampled to match the $150~\mathrm{ms}$ cadence of the former. The ion velocity was processed according to the procedure described in Sec~\ref{sec:filter}. So for the present context, Eq.~\ref{eq:3rd} can be written as
\begin{eqnarray}
Y^{\pm}(\tau) = (4/3)\epsilon^{\pm} V_{\mathrm{SW}} \tau\label{eq:3rdsw},
\end{eqnarray}
where now $Y^{\pm}(\tau) = \langle |\Delta \mathbf{Z^{\pm}}(\tau;t)|^2 \Delta Z_{R}^{\mp}(\tau;t) \rangle_{t}$. Note that the ensemble average in Eq.~\ref{eq:3rd} is now replaced by a time average $\langle \cdots \rangle_{t}$. Henceforth, unless explicitly mentioned, we shall use $\langle \cdots \rangle$ to represent time average over the total dataset interval. In Figure \ref{fig:3rd}, 
we have plotted the absolute values of the mixed third-order structure functions.
A linear scaling is indeed observed in the inertial range and interpreted here 
according to 
$|Y^{\pm}(\tau)| = (4/3)|\epsilon^{\pm}| V_{\mathrm{SW}} \tau$. By fitting a straight line in the inertial range we obtain
\begin{eqnarray}
Y^{+}(\tau) &\simeq& 0.55 \tau,\\
\frac{4}{3} \epsilon^{+} V_{\mathrm{SW}} \tau &\simeq& 0.55 \tau,\nonumber\\
\implies~\epsilon^{+} \simeq 1.1 &\times& 10^{3} \mathrm{~J~kg^{-1}~s^{-1}}.
\end{eqnarray}
A similar calculation gives $\epsilon^{-} \simeq 0.9 \times 10^{3} \mathrm{~J~kg^{-1}~s^{-1}}$. These values are comparable to the values reported in the literature for solar wind~\citep{Sorriso-Valvo2007PRL,MacBride2008ApJ}.

Next, we compute higher-order turbulence statistics. We estimate single-spacecraft values using the Taylor frozen-in hypothesis, as well as multi-spacecraft increments which allow us to estimate the values at the spacecraft separation independently of the Taylor hypothesis. The structure functions defined earlier can be generalized to higher orders as
\begin{eqnarray}
D^{(p)}_{\mathrm{R}}(l) = \langle [\Delta V^{t}_{\mathrm{R}}(l)]^{p} \rangle \label{eq:struc1sc},
\end{eqnarray}
for each single-spacecraft, Taylor-hypothesis based estimate, and
\begin{eqnarray}
D^{(p)}_{\mathrm{R}}(r_{\mathrm{ab}}) = \langle [\Delta V^{ab}_{\mathrm{R}}(t)]^{p} \rangle \label{eq:struc2sc},
\end{eqnarray}
for the two-spacecraft estimate. Here $l$ is the spatial lag computed from temporal lag using Taylor hypothesis.
\begin{figure}
\begin{center}
	\includegraphics[scale=0.6]{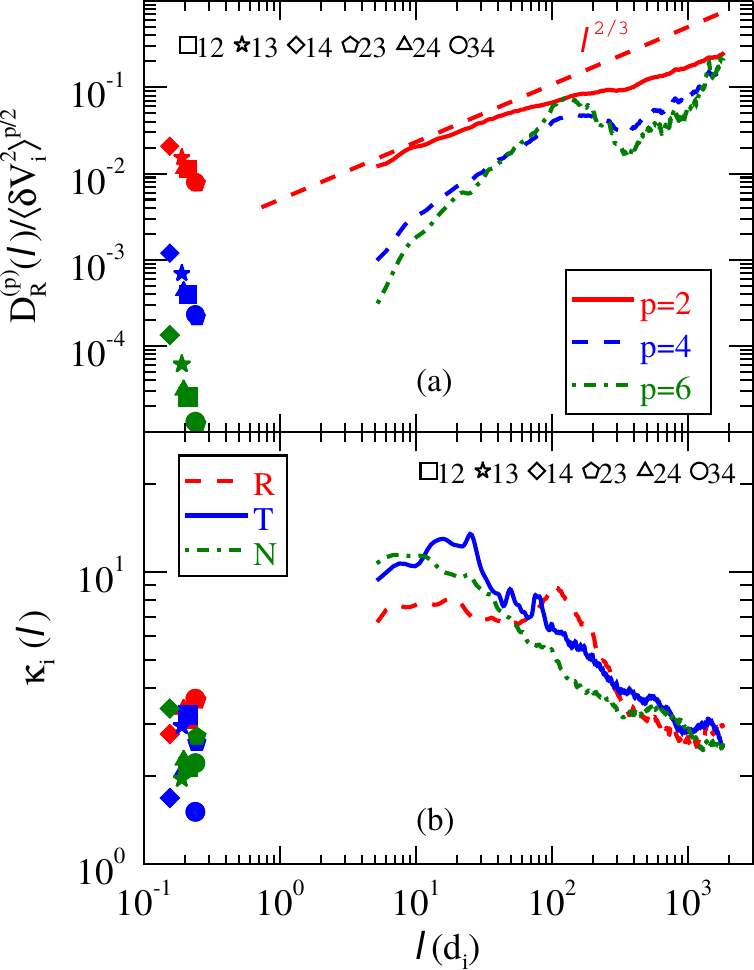}
	\caption{Panel (a): normalized ion velocity (radial) structure functions of different orders. A $l^{2/3}$ scaling is shown for reference. Panel (b): scale-dependent kurtosis for all three components of ion velocity. The symbols represent different spacecraft pairs used for the multi-spacecraft points.}
	\label{fig:struc_sdk}
\end{center}
\end{figure}
We show the second, fourth, and sixth-order structure functions for the radial $(R)$ component of the ion velocity in panel (a) of Figure~\ref{fig:struc_sdk}. The $p^\mathrm{th}$ order structure functions are normalized by the $p^\mathrm{th}$ power of the rms velocity fluctuation. The lines are single-spacecraft, Taylor hypothesis based calculations and the solid symbols represent multi-spacecraft evaluations, computed at the spatial lags given by the corresponding spacecraft separation scales. Second, fourth, and sixth-order results are plotted in color red, blue, and green respectively.

The kurtosis, or normalized fourth-order moment quantifies the deviation from 
gaussianity, and therefore provides insights into the intermittency or patchiness of the turbulence~\citep{Matthaeus2015PTRSA}. For a
Gaussian process the kurtosis is equal to three. The scale-dependent kurtosis (SDK) can be calculated using single-spacecraft time series as 
\begin{eqnarray}
\kappa_i(l) = \frac{D^{(4)}_{i}(l)}{[D^{(2)}_{i}(l)]^2}\label{sdk1sc},
\end{eqnarray}
where $i$ is the component in GSE coordinate. Similarly, SDK can be computed from multi-spacecraft measurements as
\begin{eqnarray}
\kappa_i(r_{\mathrm{ab}}) = \frac{D^{(4)}_{i}(r_{\mathrm{ab}})}{[D^{(2)}_{i}(r_{\mathrm{ab}})]^2}\label{sdk2sc}.
\end{eqnarray}
We plot the SDK at varying lag for all three components of the ion velocity in panel (b) of  Figure~\ref{fig:struc_sdk}. For SDK, the radial $(R)$, tangential $(T)$, and normal $(N)$  components are plotted in red, blue, and green respectively. The lines represent single-spacecraft estimates while the symbols are evaluated using multi-spacecraft technique.
   
It is important to note that the kurtosis reaches a peak at scales of a few $d_\mathrm{i}$. Similar results were obtained for magnetic field increments by~\cite{Chasapis2017ApJL}. Therefore, the presence of intermittent magnetic field structures at scales of a few $d_\mathrm{i}$ coincides with intermittency in ion velocity at the same scale. This suggests that such structures have an important contribution 
to kinetic effects, including 
heating, temperature anisotropies, and production of 
suprathermal particles in solar wind turbulence, consistent with previous studies~\citep{Servidio2012PRL,Greco2012PRE,Karimabadi2013PoP,Parashar2016ApJ}. 

Previous studies have observed ion-scale current sheets in the solar wind, and have 
reported particle energization 
due to magnetic reconnection in the form of reconnection outflow jets~\citep{Gosling2005JGR,Gosling2007ApJL,Gosling2012SSR,Osman2014PRL}. 
Such case studies of individual events would be consistent with the observed increase of 
intermittency of the ion velocity at exactly those scales. 
Regardless of the underlying mechanisms producing the energization
suggested by the increased intermittency of the ion velocity, 
this observation suggests that intermittent magnetic field ion-scale structures, 
such as current sheets, 
play a dominant role 
in dissipation of turbulent energy in the solar wind.

\section{Discussion}\label{sec:disc}
In this paper we have provided a first look at turbulence statistics 
using MMS FPI data at scales ranging from ``energy containing" scales down to
proton kinetic scales. This is made possible by the 
development and implementation of a time-series analysis 
procedure based on a spectral domain Hampel filter followed 
by a phase-preserving time series reconstruction. 
Specific results include first 
a demonstration 
that the electron and proton fluid velocity spectra 
track very well the magnetic spectrum down to the ion kinetic scales. However we find, from  multi-spacecraft estimates, that the fluctuation velocity of electrons exceed that of ions, which further exceed the fluctuations of magnetic field at the kinetic scales.  
We then showed an application of the Politano-Pouquet third-order law, based 
on Elsasser variables (which simultaneously 
require both magnetic field and velocity data),
also extending to kinetic scales. At small scales the mixed, third-order structure functions deviate from a linear scaling, clearly indicating the prominence of kinetic effects at those scales.  
Finally we explore the higher order statistics for 
the solar wind proton velocities, up to sixth-order increments, 
finding good qualitative agreement with previously 
published results based only on the magnetic field data by ~\cite{Chasapis2017ApJL}.
  
We note that, as far as we are aware, these are the 
first such studies of turbulence statistics involving both 
velocity and magnetic field measurements at these scales in the solar wind.
The results of the spectra 
suggests ion and electron velocity 
cascade follows the magnetic field spectral cascade
to the vicinity of the ion inertial length, as expected from 
various theoretical studies~\citep{Sonnerup1979Book,Karimabadi2013PoP}, and inferred by observations that 
make use of electric field data~\citep{Bale2005PRL}. Previously, \cite{Safrankova2013PRL,Safrankova2015ApJ,Safrankova2016ApJ} reported spectra of ion distribution moments extending to kinetic scales using data from the Faraday cups onboard Spektr-R spacecraft, but no magnetic field measurements were available. A systematic comparison would be premature at this stage since the Faraday cups and the FPI instrument operate in fundamentally different ways. Further, we consider only one selected solar wind interval in this paper. We plan to take up a detailed study of a large sample of MMS solar wind data in future. The scale-dependent kurtosis
of ion velocity shows
a peak near $l \sim d_\mathrm{i}$, suggesting
enhanced intermittency at those scales. This result indicates the presence of kinetic processes 
and particle energization that are associated with the coherent structures~\citep{Tessein2013ApJL,Chasapis2018ApJL}.
\section*{Appendix: Evaluation of the Filter}\label{sec:appendix}
\begin{figure}
	\begin{center}
	\includegraphics[scale=0.85]{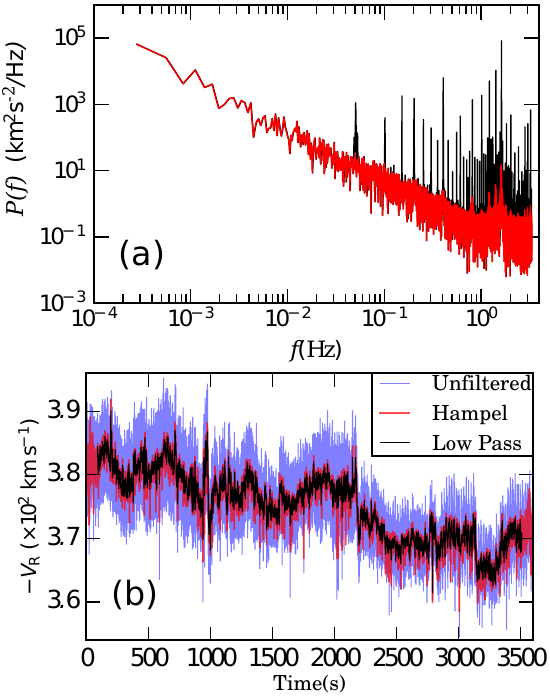}
	\caption{Demonstration of the effect of the Hampel filter on ion velocity. Panel (a): Spectrum of the unfiltered level 2 burst-mode data is shown in black. The plot in red is obtained after applying the Hampel filter on the spectrum. Panel(b): Reconstructed time series at different stages of the filter procedure (only (negative) $R$ (radial) component of the ion velocity shown here). Blue: Unfiltered level 2 burst mode data; red: after applying the Hampel filter; black: low-pass cut-off at $1~\mathrm{Hz}$ on the Hampel-filtered signal. The first and last 500 points of the black line have been omitted to indicate that those points were removed in the final analysis.}
	\label{fig:demo}
	\end{center}
\end{figure}

As described in Section~\ref{sec:filter}, for solar wind FPI moments, a main feature is the presence of large-amplitude spikes at specific frequencies. These features are presumed to be instrumental in origin and probably arises from a combination of factors. Figure~\ref{fig:demo} demonstrates the effect of applying the Hampel filter to the ion velocity spectrum in frequency space.
Panel (a) shows the original spectrum in black (as in Figure~\ref{fig:Ef_vi}), superimposed on the spectrum corresponding to the Hampel-filtered signal in red. Most of the narrow spikes are eliminated except the broad peak near $\approx 1.625~\mathrm{Hz}$. The low-frequency part of the spectrum remains immune to the filter, as desired. Panel (b) shows the real space time series (only the radial component shown here) before this procedure 
and after filtering and reconstruction of the time series. 
It is evident that the filter reduces the fluctuation amplitude of the signal by a considerable amount, leaving the global mean unchanged. The effect of presence of these spikes in the spectra becomes prominent in various standard quantities related to turbulence, e.g., the correlation function and SDK. SDK has been defined in the main text. The auto-correlation function is defined as,
\begin{eqnarray}
R(\tau) = \langle \mathbf{V}(t+\tau) \cdot \mathbf{V}(t) \rangle \label{eq:corr}.
\end{eqnarray} 
We use the correlation functions and the SDK as diagnostic tools to check if the signal is sufficiently ``clean" at different stages. 
\begin{figure*}
	\begin{center}
		\includegraphics[scale=0.5]{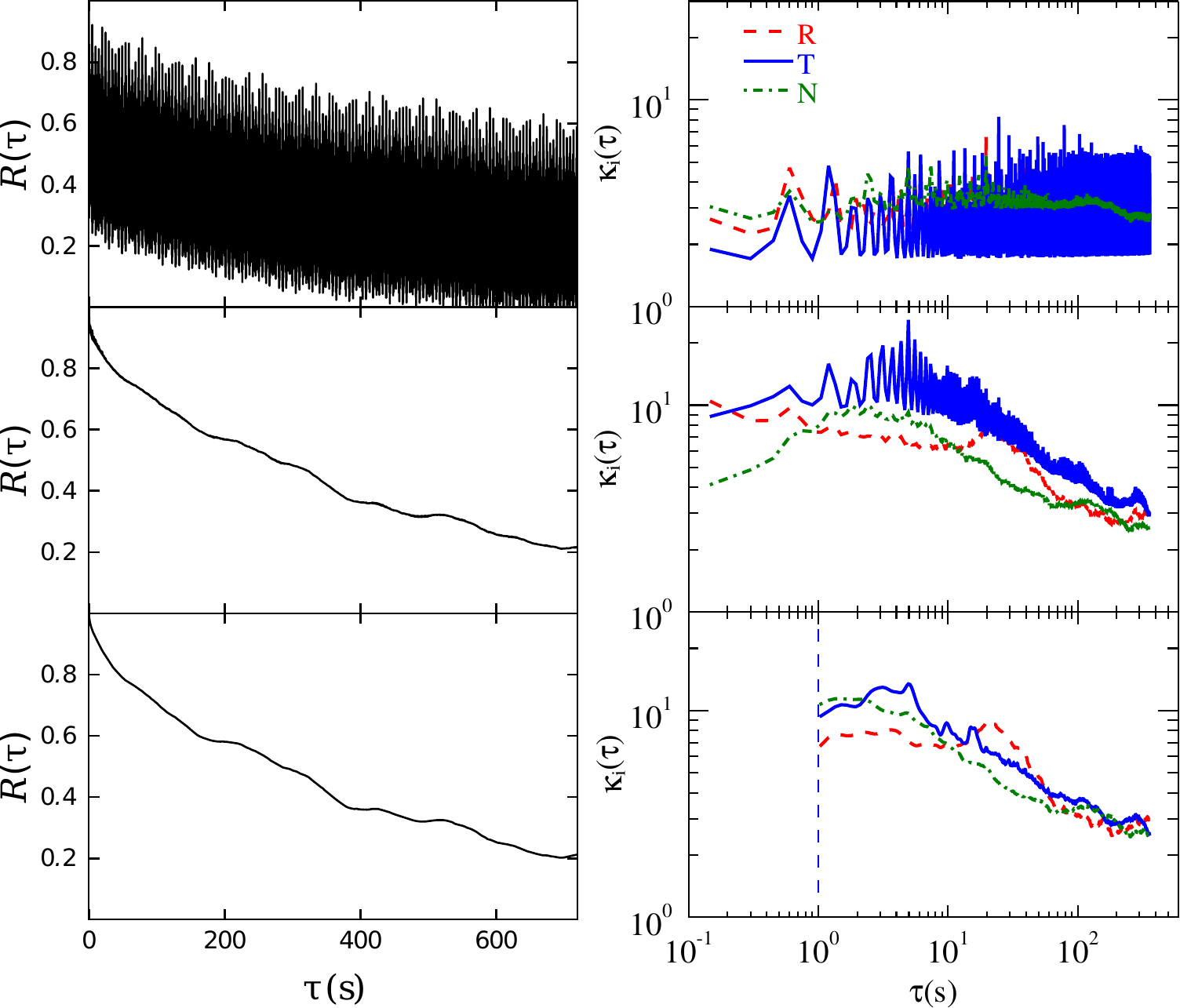}
		\caption{Normalized correlation functions and scale-dependent kurtosis for ion velocity at different stages of the filter procedure. Top panel: unfiltered signal, middle panel: Hampel-filtered, bottom panel: low-pass filtered at the noise floor $(1~\mathrm{Hz})$, and then  removing the ``edge-effects".}
		\label{fig:diagnos}
	\end{center}
\end{figure*}
Figure~\ref{fig:diagnos} shows these quantities at different stages of the procedure. As can be seen from the top panels of Figure~\ref{fig:diagnos}, the unfiltered signal is 
dominated by the spurious signals
which obscure the physical content of the signal, 
rendering it difficult to
be used for statistical studies.  
After applying the Hampel filter, however, 
the middle panels show dramatic improvement in the quality of the signal. 
As discussed in Section~\ref{sec:data}, 
we further remove the high-frequency region of the signal, 
dominated by counting statistics noise, 
by applying a sharp low-pass filter at $1~\mathrm{Hz}$. 
Finally, 
after removing the ``edges" to eliminate artificial oscillations, 
the bottom panels of Figure~\ref{fig:diagnos} appear to be free of most of the artifacts 
and usable for studies of turbulence statistics, as the results in the main text demonstrate.
\section*{Acknowledgments}
This research was partially supported by the MMS mission
through NASA grant NNX14AC39G at the University of
Delaware, by NASA LWS grant NNX17AB79G, and by the 
Parker Solar Probe Plus project through Princeton/ISOIS 
subcontract SUB0000165, and in part by NSF-SHINE AGS-1460130. A.C. is supported by the
NASA grants. W.H.M. is a member of the MMS Theory and
Modeling team. The French involvement (SCM) on MMS is supported by CNES and CNRS. We are grateful to the MMS instrument teams,
especially SDC, FPI, and FIELDS, for cooperation and
collaboration in preparing the data. The data used in this
analysis are Level 2 FIELDS and FPI data products, in
cooperation with the instrument teams and in accordance their
guidelines. All MMS data are available at \url{ https://lasp.colorado.edu/mms/sdc/}.


\end{document}